\documentclass[preprint,showpacs,preprintnumbers,amsmath,amssymb]{revtex4}

\usepackage{graphicx}
\usepackage{dcolumn}
\usepackage{bm}
\def\E^#1{{\buildrel #1 \over\vee}}

\begin{document}

\title{QUANTUM KINETIC EQUATIONS \\AND EVOLUTION OF MANY-PARTICLE SYSTEMS}%

\author{V.I. Gerasimenko}%
\email{gerasym@imath.kiev.ua}
\affiliation{Institute of Mathematics NASU,\\3 Tereshchenkivs'ka str., \\01601 Kyiv, Ukraine}

\begin{abstract}
In the paper we discuss possible approaches to the problem of the rigorous derivation of quantum kinetic equations
from underlying many-particle dynamics.
For the description of a many-particle evolution we construct solutions of the Cauchy problems of
the BBGKY hierarchy and the dual BBGKY hierarchy in suitable Banach spaces.
In the framework of the conventional approach to the description of kinetic evolution
the mean-field asymptotics of the quantum BBGKY hierarchy solution is constructed.
We develop also alternative approaches.
One method is based on the construction of the solution asymptotics of the initial-value problem of the
quantum dual BBGKY hierarchy.
One more approach is based on the generalized quantum kinetic equation that is a consequence of the
equivalence of the Cauchy problems of such evolution equation and the BBGKY hierarchy
with initial data determined by the one-particle density operator.

\vspace{2pc}
\end{abstract}

\bigskip

\maketitle

\vphantom{math}
\vskip0.7cm

\protect\tableofcontents

\newpage
\vphantom{math}
\vskip2.7cm

\section{Introduction}
We develop a formalism suggested by Bogolyubov \cite{BC},\cite{BQ} for the description
of the evolution of infinitely many particles.
The evolution equations of quantum many-particle systems arise in many
problems of modern statistical mechanics
\cite{CGP97}. In the theory of such equations during the last decade, many new results have been obtained,
in particular concerning the fundamental problem of the
rigorous derivation of quantum kinetic equations \cite{BCEP3,FL,Sp07,AA,ESchY8,AN,PP,Sh} and, among them, the kinetic equations
describing the Bose condensate \cite{BSF},\cite{BGGM1,ESchY2,M1,M2,AGT},\cite{FL},\cite{ESchY8},\cite{Sh}.

A description of quantum many-particle systems is formulated in
terms of two sets of objects: observables and states. The mean value
functional defines a duality between observables and states. As a
consequence, there exist two approaches to the description of the
evolution. Usually, the evolution of many-particle systems is
described in the framework of the evolution of states by the BBGKY
hierarchy for marginal density operators \cite{BC},\cite{BQ}, \cite{Pe95,CGP97,GPM}.
An equivalent approach to the description of the evolution of many-particle systems
is given by the dual BBGKY hierarchy \cite{BG},\cite{BG1},\cite{CGP97}
in the framework of the evolution of marginal observables.

The aim of this work is to
consider links between the many-particle quantum dynamics and quantum kinetic equations.

A conventional approach to the problem of the rigorous derivation of kinetic equations from
underlying many-particle dynamics consists in the construction of
a suitable scaling limit \cite{Gd}, for instance, the Boltzmann-Grad limit or the mean-field limit \cite{CGP97},\cite{GP90}
of a solution of the initial-value problem of the
BBGKY hierarchy. As a result, the solution limit is governed by the limit hierarchy preserving the chaos property,
and the one-particle density operator satisfies the kinetic equation \cite{BGGM1,ESchY2,M1,AGT}.
In the paper we formulate new approaches to the solving the mentioned problem which
are based on the description of a many-particle evolution
by the dual BBGKY hierarchy and by the generalized quantum kinetic equation.

We outline the structure of the paper and the main results.

In the beginning in Section II we introduce some preliminary definitions and
construct a solution of the Cauchy problem to the dual BBGKY hierarchy for marginal observables and
the canonical BBGKY hierarchy for marginal density operators of quantum many-particle systems.
We formulate also one more approach to the description of quantum many-particle dynamics
which is based on an equivalence
of the Cauchy problem of the BBGKY hierarchy with initial data determined by the one-particle density operator
and the corresponding initial value-problem for a generalized quantum kinetic equation.

In Section III, the results obtained in the previous section are used to analyze of the mean-field asymptotics
of constructed solutions, in particular to derive the nonlinear Schr\"{o}dinger equation and its generalizations.
We formulate also new methods of the derivation of quantum kinetic equations from
underlying many-particle dynamics. One method is based on the study of the scaling limits
of a solution of the initial-value problem of the dual BBGKY hierarchy. Another method is based
on a generalized quantum kinetic equation.

Finally in Section IV, we conclude with some
observations and perspectives for the future research.

\section{Dynamics of quantum many-particle systems}
We study possible approaches to the description of the
evolution of quantum many-particle systems, namely the Heisenberg and Schr\"{o}dinger pictures of the evolution.
We introduce hierarchies of evolution equations for marginal observables and states
and construct a solution of the Cauchy problems of these hierarchies in suitable Banach spaces.
We develop also one more approach based on the generalized quantum kinetic equation that is a consequence of the
equivalence of the Cauchy problems of such evolution equation and the BBGKY hierarchy
for certain class of initial data.

\subsection{The dual BBGKY hierarchy}
We will consider a quantum system of a non-fixed
(i.e. arbitrary but finite \cite{GP83}) number of identical (spinless)
particles obeying the Maxwell-Boltzmann statistics in the space $\mathbb{R}^{\nu}$. We will use units where
$h={2\pi\hbar}=1$ is the Planck constant and $m=1$ is the mass of particles.
The Hamiltonian of such a system $H={\bigoplus\limits}_{n=0}^{\infty}H_{n}$ is a self-adjoint operator with the domain
$\mathcal{D}(H)=\{\psi=\oplus_{n=0}^{\infty} \psi_{n}\in{\mathcal{F}_{\mathcal{H}}}\mid \psi_{n}\in\mathcal{D}
(H_{n})\in\mathcal{H}_{n},{\sum\limits}_{n}\|H_{n}\psi_{n}\|^{2}<\infty\}\subset{\mathcal{F}_{\mathcal{H}}},$ where
$\mathcal{F}_{\mathcal{H}}={\bigoplus\limits}_{n=0}^{\infty}\mathcal{H}^{\otimes n}$ is the Fock space over the Hilbert space
$\mathcal{H}$~ ($\mathcal{H}^{0}=\mathbb{C}$).
Assume $\mathcal{H}=L^{2}(\mathbb{R}^\nu)$~(the coordinate representation), then an element $\psi\in\mathcal{F}_{\mathcal{H}}
={\bigoplus\limits}_{n=0}^{\infty}L^{2}(\mathbb{R}^{\nu n})$ is a sequence of functions
$\psi=\big(\psi_0,\psi_{1}(q_1),\ldots,\psi_{n}(q_1,\ldots,q_{n}),\ldots\big)$
such that $\|\psi\|^{2}=|\psi_0|^{2}+\sum_{n=1}^{\infty}\int dq_1\ldots dq_{n}|\psi_{n}(q_1,\ldots ,q_{n})|^{2}<+\infty.$
On the subspace of infinitely
differentiable functions with compact supports $\psi_n\in L^{2}_0(\mathbb{R}^{\nu n})\subset L^{2}(\mathbb{R}^{\nu n})$
the $n$-particle Hamiltonian $H_{n}$ acts according to the formula ($H_{0}=0$)
\begin{eqnarray}\label{H}
      H_{n}\psi_n =\sum\limits_{i=1}^{n}K(i)\psi_n
      +\epsilon\sum\limits_{i<j=1}^{n}\Phi(i,j)\psi_{n}.
\end{eqnarray}
where $K(i)\psi_n = -\frac{1}{2}\Delta_{q_i}\psi_n$ is the operator of kinetic energy,
$\Phi(i,j)\psi_{n}= \Phi(|q_{i}-q_{j}|)\psi_{n}$ is the operator of a two-body interaction potential
satisfying Kato conditions, and $\epsilon>0$ is a scaling parameter.

Let a sequence $g=\big(g_0,g_{1},\ldots,$ $g_{n},\ldots \big)$ be an infinite sequence of self-adjoint bounded
operators $g_{n}$  defined on the Fock space
$\mathcal{F}_{\mathcal{H}}={\bigoplus\limits}_{n=0}^{\infty}\mathcal{H}^{\otimes n}$ over the Hilbert space
$\mathcal{H}$~ ($\mathcal{H}^{0}=\mathbb{C}$).
An operator $g_{n}$, defined on the $n$-particle
Hilbert space $\mathcal{H}_{n}=\mathcal{H}^{\otimes n}$ will be denoted by $g_{n}(1,\ldots,n)$.
For a system of identical particles obeying the
Maxwell-Boltzmann statistics, one has $ g_{n}(1,\ldots,n)=g_{n}(i_1,\ldots,i_n)$ for any permutation
of indices $\{i_{1},\ldots,i_{n}\}\in \{1,\ldots,n\}$.

Let the space $\mathfrak{L}(\mathcal{F}_\mathcal{H})$ be the space of sequences
$g=\big(g_0,g_{1},\ldots,$ $g_{n},\ldots \big)$ of
bounded operators $g_{n}$ defined on the Hilbert space
$\mathcal{H}_n$ and satisfying the symmetry property
$ g_{n}(1,\ldots,n)=g_{n}(i_1,\ldots,i_n)$, if $\{i_{1},\ldots,i_{n}\}\in \{1,\ldots,n\}$, with an operator norm.
We will also consider a more general space $\mathfrak{L}_{\gamma}(\mathcal{F}_\mathcal{H})$
with a norm
\begin{eqnarray*}
            &&\|g\|_{\mathfrak{L}_{\gamma} (\mathcal{F}_\mathcal{H})}=
            \max\limits_{n\geq 0}~ \frac{\gamma^n}{n!}~\|g_{n}\|_{\mathfrak{L}(\mathcal{H}_{n})},
\end{eqnarray*}
where $0<\gamma<1$ and  $\| . \|_{\mathfrak{L}(\mathcal{H}_{n})}$ is an operator norm.
An observable of the many-particle quantum system is a sequence of self-adjoint
operators from $\mathfrak{L}_{\gamma}(\mathcal{F}_\mathcal{H})$.

On the space $\mathfrak{L}_{\gamma}(\mathcal{F}_\mathcal{H})$
we consider the initial-value problem of
the dual BBGKY hierarchy.

The evolution of marginal observables is described by the initial-value problem
of the following hierarchy of evolution equations:
\begin{eqnarray}\label{dh}
       &&\frac{\partial}{\partial t}G_{s}(t,Y)=\big(\sum\limits_{i=1}^{s}\mathcal{N}_{0}(i)+
         \epsilon\sum\limits_{i<j=1}^{s}\mathcal{N}_{\mathrm{int}}(i,j)\big)G_{s}(t,Y)+\nonumber \\
       &&+\epsilon\sum_{j_1\neq j_{2}=1}^s
        \mathcal{N}_{\mathrm{int}}(j_1,j_{2})G_{s-1}(t,Y\backslash\{j_1\}),
\end{eqnarray}
\begin{eqnarray}\label{dhi}
       G_{s}(t)\mid_{t=0}=G_{s}(0), \quad\ s\geq 1.
\end{eqnarray}
In equations (\ref{dh}) we use notation $Y\equiv(1,\ldots,s)$.
The operators $\mathcal{N}_{0}, \mathcal{N}_{\mathrm{int}}$ are consequently defined on $\mathcal{D}(\mathcal{N}_{0})\subset\mathfrak{L}_{\gamma}(\mathcal{F}_\mathcal{H})$ as follows:
\begin{eqnarray}\label{c1}
    &&\mathcal{N}_{0}(j)g=-i \big[g,K(i)\big],\\
\label{ci}
    &&\mathcal{N}_{\mathrm{int}}(i,j)g=-i\big[g,\Phi(i,j)\big],
\end{eqnarray}
where $[\, \cdot\,,\,\cdot \,]$ is a commutator of operators.
We refer to the evolution equations (\ref{dh}) as the quantum dual BBGKY hierarchy, since the canonical BBGKY hierarchy \cite{CGP97}
for marginal density operators $F(t)$ is the dual hierarchy of evolution equations
with respect to the following bilinear form - average values of observables (mean values or expectation values of observables) \cite{BG}, \cite{GerSh2}:
\begin{eqnarray}\label{avmar}
        &&\big(G(t),F(0)\big)=
        \sum\limits_{s=0}^{\infty}\frac{1}{s!}
        \mathrm{Tr}_{\mathrm{1,\ldots,s}}~G_{s}(t)F_{s}(0).
\end{eqnarray}

If $\mathcal{H}=L^{2}(\mathbb{R}^\nu)$, the evolution equations (\ref{dh}) in terms of
kernels of the operators $G_{s}(t)$, $s\geq 1$, are given in the form of the equations
\begin{eqnarray*}
      && i\frac{\partial}{\partial t}G_{s}(t,q_1,\ldots,q_s;q'_1,\ldots,q'_s)=\\
    &&=\Big(-\frac{1}{2}\sum\limits_{i=1}^s(-\Delta_{q_i}+\Delta_{q'_i})
       +\epsilon\sum\limits_{1=i<j}^s\big(\Phi(q'_i-q'_j)- \Phi(q_i-q_j)\big)\Big)
        G_s(t,q_1,\ldots,q_s;q'_1,\ldots,q'_s)+\\
        &&+\epsilon\sum\limits_{1=i\neq j}^s\big(\Phi(q'_i-q'_j)
        -\Phi(q_i-q_j)\big)G_{s-1}(t,q_1,\ldots,\E^{j},\ldots,q_s;q'_1,\ldots,\E^{j},\ldots,q'_s),
\end{eqnarray*}
where $(q_1,\ldots,\E^{j},\ldots,q_s)\equiv(q_1,\ldots,q_{j-1},q_{j+1},\ldots,q_s).$

To construct a solution of the abstract initial-value problem (\ref{dh})-(\ref{dhi})
we introduce some necessary facts.

If $g\in\mathfrak{L}(\mathcal{F}_\mathcal{H})$, we introduce the group
$\mathcal{G}(t)=\oplus^{\infty}_{n=0}\mathcal{G}_{n}(t)$ of operators
\begin{eqnarray}\label{grG}
      &&\mathcal{G}_{n}(t)g_{n}\,= e^{itH_{n}}\,g_{n}\, e^{-itH_{n}}.
\end{eqnarray}
A solution of the initial-value problem
of the Heisenberg equation for observables of quantum many-particle systems is defined by this group of operators.

On the space $\mathfrak{L}_{\gamma}(\mathcal{F}_\mathcal{H})$, the one-parameter mapping
$\mathbb{R}^1\ni t\mapsto\mathcal{G}(t)g$
defines an isometric $\ast$-weak continuous group of operators, i.e. it is a $C_{0}^{\ast}$-group.
The infinitesimal generator  $\mathcal{N}={\bigoplus\limits}_{n=0}^{\infty}~
\mathcal{N}_{n}$ of the group of operators (\ref{grG}) is a closed operator for the $\ast$-weak topology
and on its domain of the definition $\mathcal{D}(\mathcal{N})\subset\mathfrak{L}_{\gamma}(\mathcal{F}_\mathcal{H})$,
which is everywhere dense for the $\ast$-weak topology, $\mathcal{N}$  is
defined in the sense of the $\ast$-weak convergence of the space
$\mathfrak{L}_{\gamma}(\mathcal{F}_\mathcal{H})$ as follows:
\begin{eqnarray}\label{infOper1}
 &&\mathrm{w^{\ast}-}\lim\limits_{t\rightarrow 0}\frac{1}{t}\big(\mathcal{G}(t)g-g\big)=i(Hg-gH)\equiv\mathcal{N}g.
\end{eqnarray}
Here, $H={\bigoplus\limits}^{\infty}_{n=0}H_{n}$ is Hamiltonian (\ref{H}) of the many-particle system
and the operator: $\mathcal{N}g=-i(gH-Hg)$ is defined on the domain $\mathcal{D}(H)\subset\mathcal{F}_\mathcal{H}.$
We remark that operator (\ref{infOper1}) is the generator of the Heisenberg equation.

We introduce the $nth$-order ($n \geq 1$) cumulant
of the groups of operators (\ref{grG}) \cite{GerR},\cite{GerRS},\cite{GerSh1}
\begin{eqnarray}\label{cumd}
\mathfrak{A}_{n}(t)\equiv\mathfrak{A}_{n}(t,X)
= \sum\limits_{\mathrm{P}:\, X ={\bigcup}_iX_i}(-1)^{|\mathrm{P}|-1}(|\mathrm{P}|-1)!
        \prod_{X_i\subset \mathrm{P}}\mathcal{G}_{|X_i|}(t),
\end{eqnarray}
where ${\sum\limits}_\mathrm{P}$
is the sum over all possible partitions $\mathrm{P} $ of the set $X\equiv(1,\ldots,n)$ into
$|\mathrm{P}|$ nonempty mutually disjoint subsets $ X_i\subset \ X$.

We indicate some properties of cumulants (\ref{cumd}) of groups of operators (\ref{grG}) \cite{G09}.
If $n=1$, for
$g_{1}\in\mathcal{D}(\mathcal{N}_1)\subset\mathfrak{L}(\mathcal{H}_1),$
the generator of the first-order cumulant
in the sense of the $\ast$-weak convergence of the space $\mathfrak{L}(\mathcal{H}_1)$ is given by operator (\ref{infOper1}), i.e.
\begin{eqnarray*}
      && \mathrm{w^{\ast}-}\lim\limits_{t\rightarrow 0}\big(\frac{1}{t}\big(\mathfrak{A}_{1}(t,1)-I\big) g_{1}
       -\big(\mathcal{N}g\big)_{1}\big)=0,
\end{eqnarray*}
where the operator $\mathcal{N}$ is defined by (\ref{infOper1}) or (\ref{c1}).
In the case $n=2$  we have, in the sense of the $\ast$-weak convergence of the space $\mathfrak{L}(\mathcal{H}_2)$,
\begin{eqnarray*}
      && \mathrm{w^{\ast}-}\lim\limits_{t\rightarrow 0}\big(\frac{1}{t}\,\mathfrak{A}_{2}(t,1,2) g_{2}
       -\epsilon\big(\,\mathcal{N}_{\mathrm{int}}(1,2)\,\big)g_{2}\big)=0.
\end{eqnarray*}
If $n>2$, as a consequence that we consider a system of particles interacting by a two-body
potential (\ref{H}), it holds
\begin{eqnarray*}
      && \mathrm{w^{\ast}-}\lim\limits_{t\rightarrow 0}\frac{1}{t}\mathfrak{A}_{n}(t) g_{n}=0.
\end{eqnarray*}

We introduce also some abridged notations: $Y\equiv(1,\ldots,s)$, $X\equiv Y\backslash\{j_1,\ldots,j_{s-n}\}$,
the set $(Y\backslash X)_1$ consists of one element of $Y\backslash X=(j_1,\ldots,j_{s-n})$,
i.e. the set $\{j_1,\ldots,j_{s-n}\}$ is a connected subset of the partition $\mathrm{P}$ ($|\mathrm{P}|=1,\; |\mathrm{P}|$
denotes the number of partitions). We will also denote the set $(Y\backslash X)_1$  by the symbol $\{j_1,\ldots,j_{s-n}\}_1$.

On the space $\mathfrak{L}_{\gamma}(\mathcal{F}_\mathcal{H})$
for the abstract initial-value problem  (\ref{dh})-(\ref{dhi})
the following statement is valid.

\emph{A solution of the initial-value problem
to the quantum dual BBGKY hierarchy (\ref{dh})-(\ref{dhi}) is determined by the expansion ($s\geq 1$)}
\begin{eqnarray}\label{sdh}
    G_{s}(t,Y)=\sum_{n=0}^s\,\frac{1}{(s-n)!} \sum_{j_1\neq\ldots\neq j_{s-n}=1}^s
      \mathfrak{A}_{1+n}\big(t,(Y\backslash X)_1,X\big) \,
      G_{s-n}(0,Y\backslash X),
\end{eqnarray}
\emph{where the operator $\mathfrak{A}_{1+n}\big(t,(Y\backslash X)_1,X\big)$
is the $(1+n)th$-order cumulant (\ref{cumd}) defined by the formula }
\begin{eqnarray*}
            \mathfrak{A}_{1+n}\big(t,(Y\backslash X)_1, X\big)=
    \sum\limits_{\mathrm{P}:\,\{(Y\backslash X)_1, X\}={\bigcup}_i X_i}
    (-1)^{\mathrm{|P|}-1}({\mathrm{|P|}-1})!\prod_{X_i\subset \mathrm{P}}\mathcal{G}_{|X_i|}(t,X_i).
\end{eqnarray*}

\emph{If $G(0)\in\mathcal{D}(\mathcal{N})\subset\mathfrak{L}_{\gamma}(\mathcal{F}_\mathcal{H})$,
it is a classical solution, and, for arbitrary initial data $G(0)\in\mathfrak{L}_{\gamma}(\mathcal{F}_\mathcal{H})$,
it is a generalized solution.}

Thus, solutions of the first two equations of hierarchy (\ref{dh}) are given by the following expansions
\begin{eqnarray*}
&&G_{1}(t,1)=\mathfrak{A}_{1}(t,1)G_{1}(0,1),\\
&&G_{2}(t,1,2)=\mathfrak{A}_{1}\big(t,\{1,2\}_1\big)G_{2}(0,1,2)
+\mathfrak{A}_{2}(t,1,2)\big(G_{1}(0,1)+ G_{1}(0,2)\big),
\end{eqnarray*}
where the first-order cumulant $\mathfrak{A}_{1}\big(t,\{1,2\}_1\big)= \mathcal{G}_{2}(t,1,2)$ is defined by group (\ref{grG}).

\subsection{The BBGKY hierarchy}
The sequence $F=(I,F_{1},\ldots,F_{n},\ldots)$ defined on the Fock space
$\mathcal{F}_{\mathcal{H}}$ of self-adjoint positive density
operators $F_{n}$ ($I$ is an identity operator) describes the state of a quantum system of a non-fixed number of particles.
The marginal density operators $F_{n}$, $n\geq 1$,  whose kernels are known
as marginal or $n$-particle density matrices, defined on the $n$-particle
Hilbert space $\mathcal{H}_{n}=\mathcal{H}^{\otimes n}$, we denote by $F_{n}(1,\ldots,n)$.
For a system of identical particles described by the
Maxwell-Boltzmann statistics, one has
$ F_{n}(1,\ldots,n)=F_{n}(i_1,\ldots,i_n)$ if $\{i_{1},\ldots,i_{n}\}\in \{1,\ldots,n\}$.

We will consider states of a system that belong to the space
$\mathfrak{L}^{1}_{\alpha}(\mathcal{F}_\mathcal{H})={\bigoplus\limits}_{n=0}^{\infty}\alpha^{n}\mathfrak{L}^{1}(\mathcal{H}_{n})$
of sequences $f=\big(I,f_{1},\ldots,f_{n},\ldots\big)$ of trace class operators
$f_{n}=f_{n}(1,\ldots,n)\in\mathfrak{L}^{1}(\mathcal{H}_{n})$ satisfying the above-mentioned symmetry condition,
equipped with the trace norm
\begin{eqnarray*}
            \|f\|_{\mathfrak{L}^{1}_{\alpha}(\mathcal{F}_\mathcal{H})}=
            \sum\limits_{n=0}^{\infty}\,\alpha^n \,\mathrm{Tr}_{\mathrm{1,\ldots,n}}|f_{n}(1,\ldots,n)|,
\end{eqnarray*}
where $\mathrm{Tr}_{\mathrm{1,\ldots,n}}$ is the partial traces over $1,\ldots,n$ particles, and $\alpha>1$ is a real number.
By $\mathfrak{L}^{1}_{\alpha, 0}$, we denote the everywhere dense set
in $\mathfrak{L}^{1}_{\alpha}(\mathcal{F}_\mathcal{H})$ of finite sequences
of degenerate operators with infinitely differentiable kernels and compact supports.

On the space $\mathfrak{L}^{1}_{\alpha}(\mathcal{F}_\mathcal{H})$,
we consider the following initial-value problem  of
the quantum BBGKY hierarchy

\begin{eqnarray}\label{2-1}
     && \frac{\partial}{\partial t}F_{s}(t)=-\big(\sum\limits_{i=1}^{s}\mathcal{N}_{0}(i)+
         \epsilon\sum\limits_{i<j=1}^{s}\mathcal{N}_{\mathrm{int}}(i,j)\big)F_{s}(t)+\nonumber\\
       &&+\sum\limits_{i=1}^{s}\mathrm{Tr}_{\mathrm{s+1}}\big(-\mathcal{N}_{\mathrm{int}}(i,s+1)\big)F_{s+1}(t),
\end{eqnarray}
\begin{eqnarray}\label{2-2}
   && F_{s}(t)\mid_{t=0}=F_{s}(0),\quad s\geq 1.
\end{eqnarray}
If $f\in \mathfrak{L}^{1}_{0}(\mathcal{F}_\mathcal{H})\subset\mathcal{D}
(\mathcal{N})\subset \mathfrak{L}^{1}_{\alpha}(\mathcal{F}_\mathcal{H})$,
the operators $\mathcal{N}_{0},$\, $\mathcal{N}_{\mathrm{int}}$ are consequently defined by (\ref{c1}), (\ref{ci}).
We remark that hierarchy (\ref{2-1}) is the dual hierarchy of equations to hierarchy (\ref{dh}).

In terms of the kernels $F_{s}(t,q_{1},\ldots,q_{s};q^{'}_{1},\ldots,q^{'}_{s})$ of $s$-particle density operators
$F_{s}(t)$, i.e. marginal or $s$-particle density matrices, equations (\ref{2-1}) take
the canonical form of the quantum BBGKY hierarchy \cite{BQ}
\begin{eqnarray*}
      &&i\frac{\partial}{\partial t} F_{s}(t,q_{1},\ldots,q_{s};q^{'}_{1},\ldots,q^{'}_{s})=\\
      &&=\Big(-\frac{1}{2}\sum\limits_{i=1}^{s}(\Delta_{q_{i}}-\Delta_{q^{'}_{i}})+\sum\limits_{i<j=1}^{s}\big(\Phi(q_{i}-q_{j})-
       \Phi(q^{'}_{i}-q^{'}_{j})\big)\Big)F_{s}(t,q_{1},\ldots,q_{s};q^{'}_{1},\ldots,q^{'}_{s})+\\
    &&+\sum\limits_{i=1}^{s}\int dq_{s+1}\big(\Phi(q_{i}-q_{s+1})-
               \Phi(q^{'}_{i}-q_{s+1})\big)F_{s+1}(t,q_{1},\ldots,q_{s},q_{s+1};q^{'}_{1},\ldots,q^{'}_{s},q_{s+1}).
\end{eqnarray*}

To construct a solution of the initial-value problem (\ref{2-1})-(\ref{2-2}), we introduce
some preliminary facts.

On the space $\mathfrak{L}^{1}_{\alpha}(\mathcal{F}_\mathcal{H})$ we define the following group
$\mathcal{G}(-t)=\oplus^{\infty}_{n=0}\mathcal{G}_{n}(-t)$ of operators:
\begin{eqnarray}\label{groupG}
      \mathcal{G}_{n}(-t)f_{n}\,:= e^{-itH_{n}}\,f_{n}\, e^{itH_{n}}.
\end{eqnarray}
On the space $\mathfrak{L}^{1}_{\alpha}(\mathcal{F}_\mathcal{H})$,
mapping (\ref{groupG})$: t\rightarrow\mathcal{G}(-t)f$
is an isometric strongly continuous group which preserves the positivity and the self-adjointness of operators.
A solution of the initial-value problem of the von Neumann equation for a statistical operator is defined
by this group.

If $f\in\mathfrak{L}_{{\alpha}, 0}^{1}(\mathcal{F}_\mathcal{H})\subset\mathcal{D}(\mathcal{N})$
in the sense of the norm convergence of the space $\mathfrak{L}^{1}(\mathcal{F}_\mathcal{H})$ there exists a limit, by which
the infinitesimal generator $-\mathcal{N}=\oplus^{\infty}_{n=0}(-\mathcal{N}_{n})$ of the group of operators (\ref{groupG})
is determined as
\begin{eqnarray}\label{infOper}
    \lim\limits_{t\rightarrow 0}\frac{1}{t}\big(\mathcal{G}(-t)f-f\big)=-i(Hf-fH):=-\mathcal{N}f,
\end{eqnarray}
where $H={\bigoplus\limits}^{\infty}_{n=0}H_{n}$ is Hamiltonian (\ref{H}) and
the operator $-i(Hf-fH)$ is defined on the domain $\mathcal{D}(H)\subset\mathcal{F}_\mathcal{H}.$
We note that operator (\ref{infOper}) is the generator of the von Neumann evolution equation.

Let $X\equiv(1,\ldots,n)$. The $nth$-order cumulant \cite{GerRS},\cite{GerSh1}
of the groups of operators (\ref{groupG}) is defined as ($n \geq 1$)
\begin{eqnarray}\label{cum}
\mathfrak{A}_{n}(-t)\equiv\mathfrak{A}_{n}(-t,X)=
 \sum\limits_{\mathrm{P}:\, X ={\bigcup}_iX_i}(-1)^{|\mathrm{P}|-1}(|\mathrm{P}|-1)!
        \prod_{X_i\subset \mathrm{P}}\mathcal{G}_{|X_i|}(-t),
\end{eqnarray}
where ${\sum\limits}_\mathrm{P}$
is the sum over all possible partitions $\mathrm{P} $ of the set $\{1,\ldots,n\}$ into
$|\mathrm{P}|$ nonempty mutually disjoint subsets $ X_i\subset \ X$.

If $n=1$, for $f_{1}\in\mathfrak{L}_{0}^{1}(\mathcal{H}_{1})\subset\mathcal{D}
(\mathcal{N}_{1})\subset\mathfrak{L}^{1}(\mathcal{H}_{1})$ in the sense of the norm convergence
in $\mathfrak{L}^{1}(\mathcal{H}_{1})$, the generator of the first-order cumulant is given by operator (\ref{infOper}), i.e.
\begin{eqnarray*}
       \lim\limits_{t\rightarrow 0}\big\|\frac{1}{t}\big(\mathfrak{A}_{1}(-t,1)-I\big)f_{1}
       -\big(-\mathcal{N}f\big)_{1}\big\|_{\mathfrak{L}^{1}(\mathcal{H}_{1})}=0.
\end{eqnarray*}
In the case $n=2$, for cumulant (\ref{cum}) we have
\begin{eqnarray*}
      && \lim\limits_{t\rightarrow 0}\big\|\frac{1}{t}\mathfrak{A}_{2}(-t)f_{2}
       -\epsilon\big(-\mathcal{N}_{\mathrm{int}}(1,2)\big)f_{2}\big\|_{\mathfrak{L}^{1}(\mathcal{H}_{2})}=0.
\end{eqnarray*}
If $n>2$, as a consequence that we consider a system of particles interacting by a two-body
potential, it holds
\begin{eqnarray*}
       &&\lim\limits_{t\rightarrow 0}\big\|\frac{1}{t}\mathfrak{A}_{n}(-t) f_{n}\big\|_{\mathfrak{L}^{1}(\mathcal{H}_{n})}=0.
\end{eqnarray*}

We introduce the following notations: $Y_{\mathrm{P}}\equiv(X_{1},\ldots,X_{|\mathrm{P}|})$ is a set,
whose elements are $|\mathrm{P}|$ mutually disjoint subsets $X_{i}\subset Y\equiv(1,\ldots,s)$
of the partition $\mathrm{P}:\,Y={\bigcup}_{i=1}^{|\mathrm{P}|}X_i.$
Since $Y_{\mathrm{P}}=(X_1,\ldots,X_{|\mathrm{P}|})$ then $Y_{1}$ is the set consisting of one element $Y=(1,\ldots,s)$
of the partition $\mathrm{P}$ ($|\mathrm{P}|=1$).
To underline that the set $(1,\ldots,s)$ is connected subset (the cluster of $s$ elements)
of a partition $\mathrm{P}$ ($|\mathrm{P}|=1$), we will also denote the set $Y_{1}$  by the symbol $\{1,\ldots,s\}_1$.

On the space $\mathfrak{L}^{1}_{\alpha}(\mathcal{F}_\mathcal{H})$
for the abstract initial-value problem  (\ref{2-1})-(\ref{2-2})
the following statement is valid \cite{GerSh1}.

\emph{If $F(0)\in \mathfrak{L}^{1}_{\alpha}(\mathcal{F}_\mathcal{H})$ and $\alpha>e$,
then, for $t\in\mathbb{R}^{1}$, there
exists a unique solution of initial-value problem (\ref{2-1})-(\ref{2-2}) given by the expansion ($s\geq1$)}
\begin{eqnarray}\label{RozvBBGKY}
     F_{s}(t,Y)= \sum\limits_{n=0}^{\infty}\frac{1}{n!}\mathrm{Tr}_{\mathrm{s+1,\ldots,s+n}}\,
      \mathfrak{A}_{1+n}(-t,Y_{1},s+1,\ldots,s+n)F_{s+n}(0,X),
\end{eqnarray}
\emph{where}
\begin{eqnarray*}
\mathfrak{A}_{1+n}(-t,Y_{1},s+1,\ldots,s+n)
= \sum\limits_{\mathrm{P}:\{Y_{1},X\setminus Y\} ={\bigcup}_iX_i}(-1)^{|\mathrm{P}|-1}(|\mathrm{P}|-1)!
        \prod_{X_i\subset \mathrm{P}}\mathcal{G}_{|X_i|}(-t)
\end{eqnarray*}
\emph{is the $(1+n)th$-order cumulant (\ref{cum}) of the groups of operators (\ref{groupG}),\, ${\sum\limits}_\mathrm{P}$
is the sum over all possible partitions $\mathrm{P} $ of the set $\{Y_{1},s+1,\ldots,s+n\}$ into
$|\mathrm{P}|$ nonempty mutually disjoint subsets $ X_i\subset \{Y_{1},X\setminus Y\}$.}

\emph{For initial data $F(0)\in \mathfrak{L}^{1}_{\alpha, 0}(\mathcal{F}_\mathcal{H})$, it is a strong solution,
and, for arbitrary initial data of the space $\mathfrak{L}^{1}_{\alpha}(\mathcal{F}_\mathcal{H})$ it is a weak solution.}

The condition $ \alpha > e $ guarantees the convergence of series  (\ref{RozvBBGKY}) and implies that the mean value
of a number of particles is finite. This fact follows if we renormalize sequence (\ref{RozvBBGKY}) in such a way:
 $\widetilde{F}_{s}(t)=\langle N\rangle^{s}F_{s}(t)$.
For arbitrary
$F(0)\in\mathfrak{L}^{1}_{\alpha}(\mathcal{F}_\mathcal{H})$, the
mean value (\ref{avmar}) of the number of particles
\begin{eqnarray}\label{N_F}
\langle N\rangle(t) = \mathrm{Tr}_{\mathrm{1}} F_{1}(t,1)
\end{eqnarray}
in state (\ref{RozvBBGKY}) is finite. In fact,
\begin{eqnarray*}
| \langle N\rangle(t)| \leq
c_{\alpha}\|F(0)\|_{\mathfrak{L}_{\alpha}^{1}(\mathcal{F}_\mathcal{H})}<\infty,
\end{eqnarray*}
where $c_{\alpha}=e^{2}(1-\frac{e}{\alpha})^{-1}$ is a constant. To describe the evolution of an infinite-particle system,
we have to construct a solution of the initial-value problem (\ref{2-1})-(\ref{2-2})
in more general spaces than $\mathfrak{L}_{\alpha}^{1}(\mathcal{F}_\mathcal{H})$.
This problem will be discussed in Conclusion.

We remark that, for classical systems of particles, the first few terms of cumulant expansion (\ref{RozvBBGKY})
for the BBGKY hierarchy were obtained in \cite{Gre56},\cite{Co68}. The methods used by Green and Cohen were based
on the analogy with the Ursell-Mayer cluster expansions for equilibrium states.

A solution of the initial-value problem (\ref{2-1})-(\ref{2-2}) is usually represented
as the perturbation (iteration) series \cite{Pe95},\cite{ESchY2},\cite{M1}.
On the space $\mathfrak{L}_{\alpha}^{1}(\mathcal{F}_\mathcal{H})$, expansion (\ref{RozvBBGKY}) is equivalent to the iteration series.
Indeed, if an interaction potential is a bounded operator, then if $f_s\in\mathfrak{L}^{1}(\mathcal{H}_s)$, for group (\ref{groupG}),
an analog of the Duhamel formula holds
\begin{eqnarray}\label{iter2kum}
       && \big(\mathcal{G}_{s}(-t,1,\ldots,s)-
           \prod\limits_{l=1}^{s}\mathcal{G}_{1}(-t,l)\big)f_s=\nonumber\\
     &&=\epsilon\int\limits_{0}^{t}d\tau
           \prod\limits_{l=1}^{s}\mathcal{G}_{1}(-t+\tau,l)\big(-\sum\limits_{i<j=1}^{s}\mathcal{N}_{\mathrm{int}}(i,j)\big)
           \mathcal{G}_{s}(-\tau)f_s.
\end{eqnarray}
Then, according to the unitary property of group (\ref{groupG})
on the space $\mathfrak{L}_{\alpha}^{1}(\mathcal{F} _\mathcal{H})$, solution expansion (\ref{RozvBBGKY}) reduces
to the iteration series of BBGKY hierarchy (\ref{2-1})
\begin{eqnarray}\label{Iter}
      &&F_{s}(t,1,\ldots,s)
        = \sum\limits_{n=0}^{\infty}\epsilon^n \int\limits_0^t dt_{1}\ldots\int\limits_0^{t_{n-1}}dt_{n}\,
          \mathrm{Tr}_{s+1,\ldots,s+n}\mathcal{G}_{s}(-t+t_{1})\times \nonumber\\
       && \times\sum\limits_{i_{1}=1}^{s}
          \big(-\mathcal{N}_{\mathrm{int}}(i_{1},s+1)\big)\mathcal{G}_{s+1}(-t_{1}+t_{2})\ldots\mathcal{G}_{s+n-1}(-t_{n-1}+t_{n})\times\nonumber\\
     && \times \sum\limits_{i_{n}=1}^{s+n-1}\big(-\mathcal{N}_{\mathrm{int}}(i_{n},s+n)\big) \mathcal{G}_{s+n}(-t_{n})F_{s+n}(0,1,\ldots,s+n).
\end{eqnarray}
If $F(0)\in\mathfrak{L}_{0}^{1}(\mathcal{F}_\mathcal{H})$, this
series exists and converges for a finite time interval \cite{Pe95}, \cite{ESchY2}.

As was mentioned above, functional (\ref{avmar}) of mean values
defines a duality between marginal observables and marginal states.
If $G(t)\in\mathfrak{L}_{\gamma}(\mathcal{F}_\mathcal{H})$ and $F(0)\in \mathfrak{L}_{\alpha}^{1}(\mathcal{F}_\mathcal{H})$, then functional (\ref{avmar}) exists, provided that $\alpha=\gamma^{-1}>e$,
and the following estimate holds:
\begin{eqnarray*}
   &&   \big|\big(G(0),F(t)\big)\big|=\big|\big(G(t),F(0)\big)\big|\leq
 e^2(1-\gamma e)^{-1}\big\|G(0)\big\|_{\mathfrak{L}_{\gamma}(\mathcal{F}_\mathcal{H})}    \big\|F(0)\big\|_{\mathfrak{L}^{1}_{{\gamma}^{-1}}(\mathcal{F}_\mathcal{H})}.
\end{eqnarray*}

\subsection{The generalized quantum kinetic equation}
We consider one more approach to the description of the evolution of states of quantum many-particle systems.
Let the initial data are completely
characterized by the one-particle density operator $ F_{1}(0)$,
for example, the initial data satisfy the chaos property (Maxwell-Boltzmann statistics)
\begin{eqnarray*}
&&F^{(c)}(0)=\big(I,F_{1}(0,1),\ldots,{\prod\limits}_{i=1}^s F_{1}(0,i),\ldots\big).
\end{eqnarray*}
In that case, the initial-value problem of
BBGKY hierarchy (\ref{2-1})-(\ref{2-2}) is not a completely well-defined Cauchy problem,
because the generic initial data are not independent for every
density operator $F_{s}(t),\, s\geq 1$, of hierarchy of equations (\ref{2-1}).
Thus, it naturally yields the opportunity of
reformulating such initial-value problem as a new Cauchy problem for the
one-particle density operator, i.e. $ F_{1}(t)$, with independent initial data $F_{1}(0)$ and
explicitly defined functionals $ F_{s}\big(t,1,\ldots,s \mid F_{1}(t)\big),\, s\geq 2,$  of the solution $ F_{1}(t)$
of this Cauchy problem instead other $s$-particle density operators $F_{s}(t),\,s\geq 2$ \cite{CGP97},\cite{GP98}.

Consequently, for an initial state satisfying the chaos property, i.e. $ F^{(c)}(0)$, the state
of a many-particle system described by the sequence $F(t)=(I,F_{1}(t,1),\ldots ,F_{s}(t,1,\ldots,s),\ldots)$
of $s$-particle density operators (\ref{RozvBBGKY}) can be described by the sequence
\begin{eqnarray*}
      && F \big(t \mid F_{1}(t)\big)=\big(I,F_{1}(t,1),F_{2}\big(t,1,2\mid F_{1}(t)\big),\ldots
       , F_{s}\big(t,1,\dots,s \mid F_{1}(t)\big),\ldots \big)
\end{eqnarray*}
of the functionals stated above.

At first, we define the sequence $F\big(t \mid F_{1}(t)\big)$ of functionals.
The functionals $ F_{s}\big(t,1,\ldots,s \mid F_{1}(t)\big),\, s\geq 2,$ are represented by the expansions over products
of the one-particle density operator $F_{1}(t)$ (for particles obeying Maxwell-Boltzmann statistics)
\begin{eqnarray}\label{f}
 &&  F_{s}\big(t,1,\ldots,s \mid F_{1}(t)\big)=
   \sum _{n=0}^{\infty }\frac{1}{n!}\mathrm{Tr}_{\mathrm{s+1,\ldots,{s+n}}} \,
   \mathfrak{V}_{1+n}(t)\prod _{i=1}^{s+n} F_{1}(t,i),
\end{eqnarray}
where the evolution operators $\mathfrak{V}_{1+n}(t)\equiv \mathfrak{V}_{1+n}(t,\{1,\ldots,s\}_1,s+1,\ldots,s+n),\, n\geq0$,
are defined from the condition that expansion (\ref{f}) of the functional $ F_{s}\big(t\mid F_{1}(t)\big)$
must be equal term by term to expansion (\ref{RozvBBGKY}) of the $s$-particle density operator $F_{s}(t)$.

The low-order evolution operators $\mathfrak{V}_{1+n}(t), \,n\geq0$, have the form
\begin{eqnarray}\label{v1}
   &&\mathfrak{V}_{1}(t,Y_{1})=\widehat{\mathfrak{A}}_{1}(t,Y_{1}),\\
\label{v2}
  && \mathfrak{V}_{2}(t,Y_{1},s+1)=
  \widehat{\mathfrak{A}}_{2}(t,Y_{1},s+1)-\widehat{\mathfrak{A}}_{1}(t,Y_{1})
          \sum_{j=1}^s \widehat{\mathfrak{A}}_{2}(t,j,s+1),
\end{eqnarray}
where $\widehat{\mathfrak{A}}_{n}(t)$ is the $nth$-order cumulant (semiinvariants) of
scattering operators
\begin{eqnarray}\label{so}
    && \widehat{\mathcal{G}}_{n}(t,1,\ldots,n):=\mathcal{G}_{n}(-t,1,\ldots,n)
     \prod _{i=1}^{n}\mathcal{G}_{1}(t,i),
\end{eqnarray}
$\widehat{\mathcal{G}}_{1}(t)=I$ is the identity operator.

In terms of scattering operators (\ref{so}) evolution operators (\ref{v1}),(\ref{v2}) get the form
\begin{eqnarray*}
    && \mathfrak{V}_{1}(t,Y_{1})=\widehat{\mathcal{G}}_{s}(t,Y),\\
    &&\mathfrak{V}_{2}(t,Y_{1},s+1)= \widehat{\mathcal{G}}_{s+1}(t,Y,s+1)-
     \widehat{\mathcal{G}}_s(t,Y)
       \sum_{j=1}^s \, \widehat{\mathcal{G}}_2(t,j,s+1)+(s-1)\,\widehat{\mathcal{G}}_s(t,Y).
\end{eqnarray*}

For $F_1(0)\in\mathfrak{L}^{1}(\mathcal{H})$ the sequence $F\big(t \mid F_{1}(t)\big)$ of functionals (\ref{f}) exists
and series (\ref{f}) converges under the condition that $\| F_1(0) \|<e^{-1}$,
i.e. if the mean value of particles is finite \cite{GP98}.

We remark that expansions (\ref{f}) are an nonequilibrium analog of expansions in powers of the density of the equilibrium
marginal density operators \cite{BC},\cite{Gre56},\cite{Co68}.

We now formulate the evolution equation for the one-particle density operator $F_{1}(t)$, i.e. for the first
element of the sequence $F\big(t \mid F_{1}(t)\big)$.
If $\| F_1(0) \|<e^{-1}$, it represents by series (\ref{RozvBBGKY}) convergent in the norm
of the space $\mathfrak{L}^{1}(\mathcal{H})$
\begin{eqnarray}\label{ske}
      F_{1}(t,1)= \sum\limits_{n=0}^{\infty}\frac{1}{n!}\mathrm{Tr}_{\mathrm{2,\ldots,{1+n}}}\,
      \mathfrak{A}_{1+n}(-t,1,\ldots,n+1)\prod _{i=1}^{n+1}F_{1}(0,i),
\end{eqnarray}
where $\mathfrak{A}_{1+n}(-t)$ is the $(1+n)th$-order cumulant (\ref{cum}) of groups of operators (\ref{groupG}).
Let $F_1(0)\in\mathfrak{L}^{1}_{0}(\mathcal{H})$. Then, by differentiating series (\ref{ske}) with respect to time variable
in the sense of the norm convergence
of the space $\mathfrak{L}^{1}(\mathcal{H}_{1})$, according to properties of cumulants (\ref{cum}),
we find that the one-particle density operator $F_{1}(t)$ is governed by
the initial-value problem of the following nonlinear evolution equation (\emph{the generalized quantum kinetic equation})
\begin{eqnarray}\label{gke}
    \frac{\partial}{\partial t}F_{1}(t,1)=-\mathcal{N}_{1}(1)F_{1}(t,1)+
    \sum\limits_{n=0}^{\infty}\frac{1}{n!}
        \mathrm{Tr}_{\mathrm{2,3,\ldots,n+2}}\big(-\mathcal{N}_{\mathrm{int}}(1,2)\big)
        \mathfrak{V}_{1+n}(t)\prod _{i=1}^{n+2} F_{1}(t,i),
\end{eqnarray}
\begin{eqnarray}\label{2}
   F_1(t,1)|_{t=0}= F_1(0,1).
\end{eqnarray}
In the kinetic equation (\ref{gke}), the evolution operators
$\mathfrak{V}_{1+n}(t)\equiv \mathfrak{V}_{1+n}(t,\{1,2\}_1,3,\ldots,2+n),\, n\geq0,$
are defined as above.

For initial-value problem (\ref{gke})-(\ref{2}) the following statement holds \cite{GP98}.

\emph{ If $F_1(0)\in\mathfrak{L}^{1}_{0}(\mathcal{H})$ is a non-negative density operator, then, provided $\| F_1(0) \|<e^{-1}$,
there exists a unique strong global in time solution of the initial-value problem (\ref{gke})-(\ref{2})
which is a non-negative density operator represented by series (\ref{ske}) convergent in the norm of the space $\mathfrak{L}^{1}(\mathcal{H})$
and a weak one for arbitrary initial data $F_1(0)\in\mathfrak{L}^{1}(\mathcal{H})$.
}

As a result, the following principle of equivalence of the initial-value
problems (\ref{2-1})-(\ref{2-2}) and (\ref{gke})-(\ref{2}) is true.

\emph{If the initial data are completely defined by the trace class operators $F_1(0)$, then the Cauchy problem (\ref{2-1})-(\ref{2-2})
is equivalent to the initial-value problem (\ref{gke})-(\ref{2})
for the generalized kinetic equation and functionals $F_{s}\big(t,1,\ldots,s \mid F_{1}(t)\big),\, s\geq 2$, defined by
expansions (\ref{f}) under the condition that $\| F_1(0) \|<e^{-1}$.}

We note that this statement is valid also for more general initial data than $F^{(c)}(0)$, namely the initial data determined by
the one-particle density operator $F_1(0)$ and operators describing initial correlations.
In this case the initial correlations are a part of the coefficients of equation (\ref{gke}) and functionals (\ref{f}).

Thus, if the initial data are completely defined by the one-particle density operator, then
all possible states of infinite-particle systems at an  arbitrary moment
of time can be described within the framework of the one-particle
density operator without any approximations.

We remark that functionals (\ref{f}) are formally concerned with the corresponding functionals
of the Bogolyubov method of the derivation of kinetic equations \cite{BC}.
Indeed, functionals (\ref{f}) and  corresponding Bogolyubov functionals
coincide if the principle of weakening of correlations for functionals (\ref{f}) holds. The proof of this assertion
is similar to the proof \cite{CGP97} of an equivalence of the
BBGKY hierarchy solution (\ref{RozvBBGKY}) and iteration series (\ref{Iter}).

\section{Derivation of nonlinear Schr\"{o}dinger equation}
We consider the problem of the rigorous derivation of quantum kinetic equations from
underlying many-particle dynamics by the example of the mean-field asymptotics of above-constructed solutions
of quantum evolution equations. In subsections B and C
we formulate new approaches to the derivation of a nonlinear Schr\"{o}dinger equation.

\subsection{The mean-field limit of the BBGKY hierarchy solution}
We present the main steps of the construction of the mean-field asymptotics
of solution (\ref{RozvBBGKY}) of the initial-value problem (\ref{2-1})-(\ref{2-2}).  For that,
we introduce some preliminary facts on the asymptotic perturbation of cumulants.

If $f_{s}\in\mathfrak{L}^{1}(\mathcal{H}_{s})$, then for an arbitrary finite
time interval, there exists the following limit of strongly continuous group (\ref{groupG}):
\begin{eqnarray}\label{lemma1}
            \lim\limits_{\epsilon\rightarrow 0}\big\|\big(\mathcal{G}_{s}(-t)-
            \prod\limits_{j=1}^{s}\mathcal{G}_{1}(-t,j)\big)f_{s}
            \big\|_{\mathfrak{L}^{1}(\mathcal{H}_{s})}=0.
\end{eqnarray}
According to an analog of the Duhamel formula (\ref{iter2kum}) and (\ref{lemma1})
for the second-order cumulant $\mathfrak{A}_{2}(-t,Y_1,s+1)$, we have
\begin{eqnarray*}
    &&\lim\limits_{\epsilon\rightarrow 0}\big\|\big(\frac{1}{\epsilon}\,\mathfrak{A}_{2}(-t,Y_1,s+1)-
    \int\limits_0^tdt_{1}\prod\limits_{j=1}^{s+1}\mathcal{G}_{1}(-t+t_{1},j)\times\\
    &&\times  \big(-\sum\limits_{i=1}^{s}\mathcal{N}_{\mathrm{int}}(i,s+1)\big)
      \prod\limits_{l=1}^{s+1}\mathcal{G}_{1}(-t_{1},l) \big)
      f_{s+1}\big\|_{\mathfrak{L}^{1}(\mathcal{H}_{s+1})}=0.
\end{eqnarray*}
In general case the following equality holds:
\begin{eqnarray}\label{Duam2}
    &&\lim\limits_{\epsilon\rightarrow 0}\big\|\big(\frac{1}{\epsilon^n}\,\mathfrak{A}_{1+n}(-t)-
    \int\limits_0^tdt_{1}\ldots\int\limits_0^{t_{n-1}}dt_{n} \prod\limits_{j=1}^{s}\mathcal{G}_{1}(-t+
       t_{1},j)\sum\limits_{i_{1}=1}^{s}\big(-\mathcal{N}_{\mathrm{int}}(i_{1},s+1)\big)\times\nonumber\\
      &&\times  \prod\limits_{j_1=1}^{s+1}\mathcal{G}_{1}(-t_{1}+t_{2},j_1)\ldots
     \prod\limits_{j_{n-1}=1}^{s+n-1}\mathcal{G}_{1}(-t_{n-1}+t_{n},j_{n-1})
        \sum\limits_{i_{n}=1}^{s+n-1}\big(-\mathcal{N}_{\mathrm{int}}(i_{n},s+n)\big)\times\nonumber\\
        &&\times\prod\limits_{j_n=1}^{s+n}\mathcal{G}_{1}(-t_{n},j_n)\big)
        f_{s+n}\big\|_{\mathfrak{L}^{1}(\mathcal{H}_{s+n})}=0.
\end{eqnarray}

Thus, if, for the initial data $F_{s}(0)\in\mathfrak{L}^{1}(\mathcal{H}_{s})$,
there exists the limit $f_{s}(0)\in\mathfrak{L}^{1}(\mathcal{H}_{s})$, i.e.,
\begin{eqnarray*}
     && \lim\limits_{\epsilon\rightarrow 0}\big\| \epsilon^{s} F_{s}(0)-f_{s}(0)\big\|_{\mathfrak{L}^{1}(\mathcal{H}_{s})}=0,
\end{eqnarray*}
then, according to (\ref{Duam2}) for an arbitrary finite time interval, there exists the mean-field limit
of solution (\ref{RozvBBGKY}) of the BBGKY hierarchy
\begin{eqnarray*}
     && \lim\limits_{\epsilon\rightarrow 0} \big\|\epsilon^{s} F_{s}(t)-f_{s}(t)\big\|_{\mathfrak{L}^{1}(\mathcal{H}_{s})}=0,
\end{eqnarray*}
where $f_{s}(t)$ is given by the series
\begin{eqnarray}\label{Iter2}
      &&f_{s}(t,1,\ldots,s)=
      \sum\limits_{n=0}^{\infty}\int\limits_0^tdt_{1}\ldots\int\limits_0^{t_{n-1}}dt_{n}\mathrm{Tr}_{\mathrm{s+1,\ldots,s+n}}
        \prod\limits_{j=1}^{s}\mathcal{G}_{1}(-t+t_{1},j)\times \nonumber\\
        &&\times \sum\limits_{i_{1}=1}^{s}\big(-\mathcal{N}_{\mathrm{int}}(i_{1},s+1)\big)
        \prod\limits_{j_1=1}^{s+1}\mathcal{G}_{1}(-t_{1}+t_{2},j_1)\ldots
         \prod\limits_{j_{n-1}=1}^{s+n-1}\mathcal{G}_{1}(-t_{n-1}+t_{n},j_{n-1})\times\nonumber\\
        &&\times\sum\limits_{i_{n}=1}^{s+n-1}\big(-\mathcal{N}_{\mathrm{int}}(i_{n},s+n)\big)
        \prod\limits_{j_n=1}^{s+n}\mathcal{G}_{1}(-t_{n},j_n)f_{s+n}(0),
\end{eqnarray}
which converges for a bounded interaction potential on a finite time interval \cite{ESchY2}.

If $f(0)\in\mathfrak{L}_{0}^{1}(\mathcal{F}_\mathcal{H})$, the sequence $f(t) = (I,f_{1}(t),\ldots,$ $f_s(t),\ldots)$ of
limit marginal density operators (\ref{Iter2}) is a strong solution of the Cauchy problem
of the \emph{Vlasov hierarchy}
\begin{eqnarray}\label{BBGKYlim}
       \frac{\partial}{\partial t}f_{s}(t)=\sum\limits_{i=1}^{s}\big(-\mathcal{N}_{0}(i)\big)f_{s}(t)+
       \sum\limits_{i=1}^{s}\mathrm{Tr}_{\mathrm{s+1}}\big(-\mathcal{N}_{\mathrm{int}}(i,s+1)\big)f_{s+1}(t),
\end{eqnarray}
\begin{eqnarray}\label{BBGKYlim0}
     f_{s}(t)|_{t=0}=f_{s}(0),\qquad s\geq 1.
\end{eqnarray}

We observe that, if the initial data satisfy the chaos property (for particles obeying Maxwell-Boltzmann statistics)
\begin{eqnarray*}
f_{s}(t,1,\ldots,s)|_{t=0}=\prod \limits_{j=1}^{s}f_{1}(0,j),\quad s\geq 2,
\end{eqnarray*}
then solution (\ref{Iter2}) of the initial-value problem of the Vlasov hierarchy (\ref{BBGKYlim})-(\ref{BBGKYlim0})
possesses of the same property
\begin{eqnarray}\label{chaosv}
f_{s}(t,1,\ldots,s)=\prod\limits_{j=1}^{s}f_{1}(t,j), \quad s\geq 2.
\end{eqnarray}

To established equality (\ref{chaosv}), we introduce marginal correlation density operators \cite{GerSh2}
\begin{eqnarray}\label{Cor}
  G_s(t,1,\ldots,s)=
  \sum\limits_{n=0}^{\infty}\frac{1}{n!}\mathrm{Tr}_{\mathrm{{s+1},\ldots ,s+n}}\,
  \mathfrak{A}_{s+n}(-t,1,\ldots,s+n)\prod_{i=1}^{s+n}G_{1}(0,i),
\end{eqnarray}
where $\mathfrak{A}_{s+n}(-t)\equiv\mathfrak{A}_{s+n}(-t,1,\ldots,s+n)$
is the $(s+n)th$-order cumulant (\ref{cum}) of the groups of operators (\ref{groupG}), and  $G_{1}(0)=F_{1}(0)$.
In the same way as (\ref{Duam2}) for arbitrary $t\in\mathbb{R}$, we establish the equality
\begin{eqnarray}\label{lemma2}
            \lim\limits_{\epsilon\rightarrow0}\big\|\frac{1}{\epsilon^{n}}\,
            \mathfrak{A}_{s+n}(-t,1,\ldots,s+n)f_{s+n}\big\|_{\mathfrak{L}^{1}(\mathcal{H}_{s+n})}=0.
\end{eqnarray}

Let
\begin{eqnarray*}
      \lim\limits_{\epsilon\rightarrow 0}\big\| \epsilon \,G_{1}(0)- f_{1}(0)\big\|_{\mathfrak{L}^{1}(\mathcal{H}_{1})}=0
\end{eqnarray*}
hold. Then, according to (\ref{lemma2}) for the correlation density operators (\ref{Cor}), we obtain
\begin{eqnarray}\label{cori}
            \lim\limits_{\epsilon\rightarrow0}\big\|\epsilon^{s}\,
            G_s(t)\big\|_{\mathfrak{L}^{1}(\mathcal{H}_{s})}=0.
\end{eqnarray}

In view of the fact that the marginal density operators (\ref{RozvBBGKY})
are expressed in terms of the correlation density operators (\ref{Cor})  by the cluster expansions
\begin{eqnarray*}
  F_{s}(t,Y) = \prod_{i=1}^{s}F_{1}(t,i)+
  \sum\limits_{\mbox{\scriptsize $\begin{array}{c}
       \mathrm{P}: \{Y\}=\bigcup_{i} X_{i},\\|\mathrm{P}|\neq s
       \end{array}$}}\prod_{X_i\subset \mathrm{P}}G_{|X_i|}(t,X_i), \quad s\geq2,
\end{eqnarray*}
and taking equality (\ref{cori}) into account, the following statement is valid.

\emph{If there exists the mean-field limit of the initial data $F_{s}(0)\in\mathfrak{L}^{1}(\mathcal{H}_{s})$ }
\begin{eqnarray*}
      \lim\limits_{\epsilon\rightarrow 0}\big\| \epsilon^{s} F_{s}(0,1,\ldots,s)-\prod\limits_{j=1}^{s} f_{1}(0,j)\big\|_{\mathfrak{L}^{1}(\mathcal{H}_{s})}=0,
\end{eqnarray*}
\emph{then, for a finite time interval for solution (\ref{RozvBBGKY}) of the BBGKY hierarchy the limit}
\begin{eqnarray*}
      \lim\limits_{\epsilon\rightarrow 0} \big\|\epsilon^{s} F_{s}(t,1,\ldots,s)-\prod\limits_{j=1}^{s}
      f_{1}(t,j)\big\|_{\mathfrak{L}^{1}(\mathcal{H}_{s})}=0
\end{eqnarray*}
\emph{holds, where $f_{1}(t)$ is the solution of the Cauchy problem of the quantum Vlasov equation}
\begin{eqnarray}\label{Vlasov1}
      \frac{\partial}{\partial t}f_{1}(t,1)=\big(-\mathcal{N}_{0}(1)\big)f_{1}(t,1)+
       \mathrm{Tr}_{2}\big(-\mathcal{N}_{\mathrm{int}}(1,2)\big)f_{1}(t,1)f_{1}(t,2),
\end{eqnarray}
\begin{eqnarray}\label{Vlasov2}
              f_{1}(t)|_{t=0}=f_{1}(0).
\end{eqnarray}

Thus, in consequence of the chaos property (\ref{chaosv}), we derive the quantum Vlasov kinetic equation (\ref{Vlasov1}).

\subsection{On the nonlinear Schr\"{o}dinger equation}
For a system in the pure state, i.e.  $f_{1}(t)=|\psi_{t}\rangle\langle\psi_{t}|$ (
$P_{\psi_{t}}\equiv |\psi_{t}\rangle\langle\psi_{t}|$ is a one-dimensional projector
onto a unit vector $|\psi_{t}\rangle$) or in terms of the kernel
$f_{1}(t,q,q')=\psi(t,q)\psi(t,q')$ of the marginal one-particle density
operator $f_{1}(t)$, the Vlasov kinetic equation (\ref{Vlasov1})
is transformed to the \emph{Hartree equation}
\begin{eqnarray}\label{Hartree}
      i\frac{\partial}{\partial t} \psi(t,q) = -\frac{1}{2}\Delta_{q}\psi(t,q)+
       \int dq'\Phi(q-q')|\psi(t,q')|^{2}\psi(t,q).
\end{eqnarray}
If the kernel of the interaction potential $\Phi(q)=\delta(q)$ is the Dirac measure, then from (\ref{Hartree})
we derive the cubic \emph{nonlinear Schr\"{o}dinger equation}
\begin{eqnarray*}
      i\frac{\partial}{\partial t} \psi(t,q)=
     -\frac{1}{2}\Delta_{q}\psi(t,q)+ |\psi(t,q)|^{2}\psi(t,q).
\end{eqnarray*}

Thus, the following statement holds:
\begin{eqnarray*}
      \lim\limits_{\epsilon\rightarrow 0}\big\|\, \epsilon^{s} F_{s}(t)-
      |\psi_{t}\rangle\langle\psi_{t}|^{\otimes s}\,\big\|_{\mathfrak{L}^{1}(\mathcal{H}_{s})}=0,
\end{eqnarray*}
where $|\psi_{t}\rangle$ is the solution of the cubic nonlinear Schr\"{o}dinger equation.

In the case of representation (\ref{Iter}) of the solution of the Cauchy problem (\ref{2-1})-(\ref{2-2}) of the BBGKY hierarchy
by the iteration series the last statement is proved in works \cite{BGGM1} -\cite{AGT}.

\subsection{The mean-field limit of the generalized quantum kinetic equation}
We construct the mean-field limit of a solution of the initial-value problem of the generalized kinetic equation (\ref{gke}).

If there exists the limit $f_{1}(0)\in\mathfrak{L}^{1}(\mathcal{H}_{1})$ of initial data (\ref{2}),
\begin{eqnarray*}
      \lim\limits_{\epsilon\rightarrow 0}\big\| \epsilon \, F_{1}(0)- f_{1}(0)\big\|_{\mathfrak{L}^{1}(\mathcal{H}_{1})}=0,
\end{eqnarray*}
then, according to (\ref{lemma1}) and (\ref{Duam2}) for an arbitrary finite time interval there exists the limit
of solution (\ref{ske}) of the generalized kinetic equation (\ref{gke})
\begin{eqnarray}\label{1lim}
      \lim\limits_{\epsilon\rightarrow 0} \big\|\epsilon \,F_{1}(t)- f_{1}(t)\big\|_{\mathfrak{L}^{1}(\mathcal{H}_{1})}=0,
\end{eqnarray}
where $f_{1}(t)$ is a strong solution of the Cauchy problem (\ref{Vlasov1})-(\ref{Vlasov2})
of the quantum Vlasov equation represented in the form of the expansion
\begin{eqnarray}\label{viter}
      &&f_{1}(t,1)=\sum\limits_{n=0}^{\infty}\int\limits_0^tdt_{1}\ldots\int\limits_0^{t_{n-1}}dt_{n}\mathrm{Tr}_{\mathrm{s+1,\ldots,s+n}}
        \prod\limits_{j=1}^{s}\mathcal{G}_{1}(-t+t_{1},j)\times\nonumber\\
        &&\times \sum\limits_{i_{1}=1}^{s}\big(-\mathcal{N}_{\mathrm{int}}(i_{1},s+1)\big)
        \prod\limits_{j_1=1}^{s+1}\mathcal{G}_{1}(-t_{1}+t_{2},j_1)\ldots
        \prod\limits_{j_{n-1}=1}^{s+n-1}\mathcal{G}_{1}(-t_{n-1}+t_{n},j_{n-1})\times\nonumber\\
      && \times\sum\limits_{i_{n}=1}^{s+n-1}\big(-\mathcal{N}_{\mathrm{int}}(i_{n},s+n)\big)
        \prod\limits_{j_n=1}^{s+n}\mathcal{G}_{1}(-t_{n},j_n)\prod\limits_{i=1}^{s+n}f_{1}(0,i),
\end{eqnarray}
and the operator $\mathcal{N}_{\mathrm{int}}$ is defined by formula (\ref{ci}).
For bounded interaction potentials, series (\ref{viter}) converges for a
finite time interval.

If $f_s\in\mathfrak{L}^{1}(\mathcal{H}_s)$ and an interaction potential is a bounded operator, then
for scattering operators (\ref{so}) an analog of the Duhamel formula holds:
\begin{eqnarray}\label{dcum}
        \big(\widehat{\mathcal{G}}_{s}(t,1,\ldots,s)-I\big)f_s=
        \epsilon\int\limits_{0}^{t}d\tau
           \prod\limits_{l=1}^{s}\mathcal{G}_{1}(\tau,l)\big(-\sum\limits_{i<j=1}^{s}\mathcal{N}_{\mathrm{int}}(i,j)\big)
          \mathcal{G}_{s}(-\tau)f_s.
\end{eqnarray}
Then, according to definition (\ref{v1}) of the evolution operators
$\mathfrak{V}_{1+n}(t,\{1,\ldots,s\}_1,s+1,\ldots,s+n),\, n\geq0,$ from expansion (\ref{f})
and equality (\ref{dcum}) we establish
\begin{eqnarray*}
   \lim\limits_{\epsilon\rightarrow 0}\big\|\big(\mathfrak{V}_{1}(t,\{1,\ldots,s\}_1)-I\big)f_{s}
            \big\|_{\mathfrak{L}^{1}(\mathcal{H}_{s})}=0.
\end{eqnarray*}
Correspondingly for $n\geq1$, it holds:
\begin{eqnarray*}
   \lim\limits_{\epsilon\rightarrow 0}\big\|\mathfrak{V}_{1+n}(t)f_{s+n}
            \big\|_{\mathfrak{L}^{1}(\mathcal{H}_{s+n})}=0.
\end{eqnarray*}

Since a solution of the initial-value problem (\ref{gke})-(\ref{2}) of the generalized kinetic equation converges
to a solution of the initial-value problem (\ref{Vlasov1})-(\ref{Vlasov2}) of the quantum Vlasov kinetic equation
as (\ref{1lim}), for functionals (\ref{f}) we obtain
\begin{eqnarray*}
  \lim\limits_{\epsilon\rightarrow 0} \big\|\epsilon^{s} F_{s}\big(t,1,\ldots,s \mid F_{1}(t)\big)- \prod\limits_{j=1}^{s}
      f_{1}(t,j)\big\|_{\mathfrak{L}^{1}(\mathcal{H}_{s})}=0,
\end{eqnarray*}
where $f_{1}(t)$ is defined by series (\ref{viter}) which converges for a
finite time interval.
The last equality means that the chaos property (\ref{chaosv}) preserves in time in the mean-field scaling limit.

Thus, we conclude that the results of the previous
subsection concerning the derivation of the Hartree equation and the nonlinear Schr\"{o}dinger equation
take place also in the case of the generalized quantum kinetic equation (\ref{gke}).

\subsection{The mean-field limit of the dual BBGKY hierarchy solution}
Consider the mean-field limit of a solution of the initial-value problem of dual BBGKY hierarchy (\ref{dh}).

For an arbitrary finite
time interval, there exists the following limit of the $\ast$-weak continuous group of operators (\ref{grG})
in the sense of the $\ast$-weak convergence of the space $\mathfrak{L}(\mathcal{H}_s)$
\begin{eqnarray}\label{Kato}
 \mathrm{w^{\ast}-}\lim\limits_{t\rightarrow 0}\big(\mathcal{G}_s(t)g_s-\prod\limits_{j=1}^{s}\mathcal{G}_{1}(-t,j)g_s\big)=0.
\end{eqnarray}
According to an analog of the Duhamel formula (\ref{iter2kum}) and (\ref{Kato})
for the second-order cumulant $\mathfrak{A}_{2}(t,1,2)$  in the same sense as above, it holds:
\begin{eqnarray}\label{KatoD}
    \mathrm{w^{\ast}-}\lim\limits_{\epsilon\rightarrow 0}\big(\frac{1}{\epsilon}\,\mathfrak{A}_{2}(t,1,2)g_{2}-
    \int\limits_0^tdt_{1}\prod\limits_{j=1}^{2}\mathcal{G}_{1}(t-t_{1},j)\mathcal{N}_{\mathrm{int}}(1,2)
      \prod\limits_{l=1}^{2}\mathcal{G}_{1}(t_{1},l)g_{2} \big)=0.
\end{eqnarray}

Thus, if, for initial data $G_{s}(0)\in\mathfrak{L}(\mathcal{H}_{s})$,
there exists the limit $g_{s}(0)\in\mathfrak{L}(\mathcal{H}_{s})$, i.e. if it holds:
\begin{eqnarray}\label{asumdin}
      \mathrm{w^{\ast}-}\lim\limits_{\epsilon\rightarrow 0}\big( \epsilon^{-s} G_{s}(0)-g_{s}(0)\big)=0,
\end{eqnarray}
then, according to (\ref{Kato}) and (\ref{KatoD}) for an arbitrary finite time interval
there exists the mean-field limit of solution (\ref{sdh})
of the dual BBGKY hierarchy (\ref{dh}) in the sense of the $\ast$-weak convergence of the space
$\mathfrak{L}(\mathcal{H}_s)$:
\begin{eqnarray}\label{asymt}
      \mathrm{w^{\ast}-}\lim\limits_{\epsilon\rightarrow 0} \big(\epsilon^{-s} G_{s}(t)-g_{s}(t)\big)=0.
\end{eqnarray}
The limit operator $g_{s}(t)$ in (\ref{asymt}) is given by the expansion
\begin{eqnarray}\label{Iterd}
      &&g_{s}(t,Y)
       =\sum\limits_{n=0}^{s-1}\,\int\limits_0^tdt_{1}\ldots\int\limits_0^{t_{n-1}}dt_{n}
       \, \mathcal{G}_{s}^{0}(t-t_{1})\sum\limits_{i_{k_{1}}\neq i_{k_{2}}=1}^{s} \mathcal{N}_{\mathrm{int}}(i_{k_{1}},i_{k_{2}})
        \,\mathcal{G}_{s-1}^{0}(t_{1}-t_{2})\ldots\nonumber\\
       &&\ldots \mathcal{G}_{s-n+1}^{0}(t_{n-1}-t_{n})
       \sum\limits_{i_{k_{n}}\neq i_{k_{n+1}}=1}^{s}\mathcal{N}_{\mathrm{int}}(i_{k_{n}},i_{k_{n+1}})\mathcal{G}_{s-n}^{0}(t_{n})\,g_{s-n}(0,Y\backslash \{i_{k_{1}},\ldots,i_{k_{n}}\}),
\end{eqnarray}
where
\begin{eqnarray*}
\mathcal{G}_{s-n+1}^{0}(t_{n-1}-t_{n})\equiv\mathcal{G}_{s-n+1}^{0}(t_{n-1}-t_{n},Y \backslash \{i_{k_{1}},\ldots,i_{k_{n-1}}\})=
\prod\limits_{j\in Y \backslash \{i_{k_{1}},\ldots,i_{k_{n-1}}\}}\mathcal{G}_{1}(t_{n-1}-t_{n},j)
\end{eqnarray*}
is the group of operators (\ref{grG}) of noninteracting particles.

If $g(0)\in\mathfrak{L}(\mathcal{F}_\mathcal{H})$, the sequence $g(t)=(g_0,g_1(t),\ldots,$ $g_{s}(t),\ldots)$
of limit marginal observables (\ref{Iterd}) is a generalized solution of the initial-value problem of
the \emph{dual Vlasov hierarchy}
\begin{eqnarray}\label{vdh}
      \frac{\partial}{\partial t}g_{s}(t,Y)=\sum\limits_{i=1}^{s}\mathcal{N}_{0}(i)\,g_{s}(t,Y)+
     \sum_{j_1\neq j_{2}=1}^s\mathcal{N}_{\mathrm{int}}(j_1,j_{2})\,g_{s-1}(t,Y\backslash\{j_1\}),
\end{eqnarray}
\begin{eqnarray}\label{vdhi}
       g_{s}(t)\mid_{t=0}=g_{s}(0), \quad\!\quad\!\! s\geq 1.
\end{eqnarray}

Consider the mean-field limit of the additive-type observables, i.e.
\begin{eqnarray*}
       G^{(1)}(0)=(0,G_{1}^{(1)}(0,1),0,\ldots).
\end{eqnarray*}
In that case, solution (\ref{sdh}) of the dual BBGKY hierarchy (\ref{dh}) has the form
\begin{eqnarray}\label{sad}
       G_{s}^{(1)}(t,Y)=\mathfrak{A}_{s}(t,Y)\,\sum_{j=1}^s\,G_{1}^{(1)}(0,j).
\end{eqnarray}

If, for the additive-type observables $G^{(1)}(0)$, condition (\ref{asumdin}) holds, i.e.
\begin{eqnarray*}
 \mathrm{w^{\ast}-}\lim\limits_{\epsilon\rightarrow 0}\big( \epsilon^{-1} G_{1}^{(1)}(0)-g_{1}^{(1)}(0)\big)=0,
\end{eqnarray*}
then, according to statement (\ref{asymt}), for (\ref{sad}) we have
\begin{eqnarray*}
      \mathrm{w^{\ast}-}\lim\limits_{\epsilon\rightarrow 0} \big(\epsilon^{-s} G_{s}^{(1)}(t)-g_{s}^{(1)}(t)\big)=0,
\end{eqnarray*}
where
\begin{eqnarray*}
  &&g_{1}^{(1)}(t,1)=\mathcal{G}_{1}(t,1)g_{1}^{(1)}(0,1),\\
   &&g_{2}^{(1)}(t,1,2)=
   \int\limits_0^t dt_{1}\prod\limits_{j=1}^{2}\mathcal{G}_{1}(t-t_{1},j)\mathcal{N}_{\mathrm{int}}(1,2)
      \sum\limits_{l=1}^{2}\mathcal{G}_{1}(t_{1},l)g_{1}^{(1)}(0,l)
\end{eqnarray*}
or as a special case of (\ref{Iterd}) the limit operator $g_{s}^{(1)}(t)$ is defined by the expansion
\begin{eqnarray}\label{itvad}
      && g_{s}^{(1)}(t,Y)=\int\limits_0^t dt_{1}\ldots\int\limits_0^{t_{s-2}}dt_{s-1}
        \, \mathcal{G}_{s}^{0}(t-t_{1})\sum\limits_{i_{k_{1}}\neq i_{k_{2}}=1}^{s} \mathcal{N}_{\mathrm{int}}(i_{k_{1}},i_{k_{2}})
        \,\mathcal{G}_{s-1}^{0}(t_{1}-t_{2})\ldots\nonumber\\
      &&\ldots \mathcal{G}_{2}^{0}(t_{s-2}-t_{s-1})
       \sum\limits_{i_{k_{s-1}}\neq i_{k_{s}}=1}^{s}\mathcal{N}_{\mathrm{int}}(i_{k_{s-1}},i_{k_{s}})
       \mathcal{G}_{1}^{0}(t_{s-1})\,g_{1}^{(1)}(0,Y\backslash \{i_{k_{1}},\ldots,i_{k_{s-1}}\}).
\end{eqnarray}

Let the initial state satisfy the chaos property (\ref{chaosv})
\begin{eqnarray*}
f_{s}^{(c)}(0,1,\ldots,s)=\prod \limits_{j=1}^{s}f_{1}(0,j),\quad s\geq 2.
\end{eqnarray*}
Then, if $g(t)\in\mathfrak{L}_{\gamma}(\mathcal{F}_\mathcal{H})$ and $f_1(0)\in \mathfrak{L}^{1}(\mathcal{H}_{1})$,
the mean value functional
\begin{eqnarray*}
       \big(g(t),f(0)\big)=
        \sum\limits_{s=0}^{\infty}\,\frac{1}{s!}\,
        \mathrm{Tr}_{\mathrm{1,\ldots,s}}\,g_{s}(t,1,\ldots,s)\prod \limits_{i=1}^{s} f_{1}(0,i)
\end{eqnarray*}
exists, provided that $\|f_1(0)\|_{\mathfrak{L}^{1}(\mathcal{H}_{1})}<\gamma$.

In consequence of the equality
\begin{eqnarray*}
       \big(g^{(1)}(t),f^{(c)}(0)\big)=
        \sum\limits_{s=0}^{\infty}\,\frac{1}{s!}\,
        \mathrm{Tr}_{\mathrm{1,\ldots,s}}\,g_{s}^{(1)}(t)\prod \limits_{i=1}^{s} f_{1}(0,i)
   = \mathrm{Tr}_{\mathrm{1}}\,g_{1}^{(1)}(0)f_{1}(t,1),
\end{eqnarray*}
where $g_{s}^{(1)}(t)$ is given by (\ref{itvad}) and $f_{1}(t,1)$ is solution (\ref{viter})
of the quantum Vlasov equation (\ref{Vlasov1}),
we find that the initial-value problem (\ref{vdh})-(\ref{vdhi}) for additive-type observables and the initial state $f^{(c)}(0) $
describes the evolution of quantum many-particle systems as by the Vlasov kinetic equation.

Correspondingly, the chaos property (\ref{chaosv}) in the Heisenberg picture of evolution of quantum many-particle systems
is fulfilled, which follows from the equality ($k\geq2$)
\begin{eqnarray*}
       \big(g^{(k)}(t),f^{(c)}(0)\big)
        =\sum\limits_{s=0}^{\infty}\,\frac{1}{s!}\,
        \mathrm{Tr}_{\mathrm{1,\ldots,s}}\,g_{s}^{(k)}(t)\prod \limits_{i=1}^{s} f_{1}(0,i)=
      \frac{1}{k!}\mathrm{Tr}_{\mathrm{1,\ldots,k}}\,g_{k}^{(k)}(0)\prod \limits_{i=1}^{k} f_{1}(t,i),
\end{eqnarray*}
where $f_{1}(t,1)$ is given by expansion (\ref{viter}).

Thus, if the initial state is a pure state, i.e.  $f_{s}(0)=|\psi_{0}\rangle\langle\psi_{0}|^{\otimes s}$,
we conclude that in the Heisenberg picture of evolution, the initial-value problem (\ref{vdh})-(\ref{vdhi}) of the dual Vlasov hierarchy
describes the evolution of quantum many-particle systems which is governed by the Hartree equation (\ref{Hartree})
in the Schr\"{o}dinger picture of evolution or it is governed by the
cubic nonlinear Schr\"{o}dinger equation, if the interaction potential $\Phi(q)=\delta(q)$ is the Dirac measure.

\section{Conclusion}
The concept of cumulants (\ref{cumd}) of the groups of operators (\ref{grG}) of the Heisenberg equations
or cumulants (\ref{cum}) of groups of operators (\ref{groupG}) of the von Neumann equations
forms the basis of the groups of operators for quantum evolution equations as well as the quantum dual BBGKY hierarchy and
the BBGKY hierarchy for marginal density operators \cite{GerRS},\cite{GerSh1},\cite{G09}.

As was mention above, for the initial data
$F(0)\in\mathfrak{L}^{1}_{\alpha}(\mathcal{F}_\mathcal{H})$,  the
average number (\ref{N_F}) of particles is finite.
In order to describe the evolution of infinitely many particles \cite{CGP97},
we have to construct solutions for initial marginal density operators belonging to more general Banach spaces
than  $\mathfrak{L}^{1}_{\alpha}(\mathcal{F}_\mathcal{H})$.
For example, it can be the space of sequences of bounded operators containing the equilibrium states \cite{Pe95}.
In that case, every term of solution expansions for BBGKY hierarchy (\ref{2-1}) and correspondingly
for the generalized quantum kinetic equation (\ref{gke}) and functionals (\ref{f}) contains the divergent traces \cite{CGP97},\cite{Co68},
In the case of the dual BBGKY hierarchy (\ref{dh}),
the problem consists in the definition of mean value functional (\ref{avmar})
which contains the divergent traces \cite{BG},\cite{G09}. The analysis of such a question
for quantum systems remains an open problem.

We have formulated two new approaches to the rigorous derivation of quantum kinetic equations from
underlying many-particle dynamics. These approaches enable one to describe the
kinetic evolution if the chaos property (\ref{chaosv}) is not fulfilled initially, i.e. in the presence
of initial correlations. Such Cauchy problem takes place in the case of kinetic evolution of the Bose condensate \cite{BQ},\cite{BSF}.
As a result of these approaches, we can formulate the kinetic equations both for a Bose gas and a Bose condensate, i.e.
the nonlinear Schr\"{o}dinger equation and the Gross–-Pitaevskii equation \cite{ESchY2},\cite{M1}, respectively.

\end{document}